\documentstyle[12pt]{article}
\title{\bf Quantization of Two-Dimensional Gravity
       with Dynamical Torsion}
\author{P. M. Lavrov\thanks{E-mail: lavrov@tspi.tomsk.su},
       P. Yu. Moshin\\
       {\it Tomsk State Pedagogical University, Tomsk 634041, Russia}}
\date{}
\begin{document}
\maketitle
  \begin{quotation}
  \small\noindent
  We consider two-dimensional gravity with dynamical torsion in the BV
  and BLT formalisms of gauge theories quantization as well as in the
  background field method.
\end{quotation}
\section{Introduction}
 Two-dimensional theories of gravity$^{1-9}$ and superagrvity$^{10-12}$ are
 interesting mainly due to their close relation to string and superstring
 theory. On the other hand, simple two-dimensional models provide a deeper
 insight into classical and quantum properties of gravity in higher
 dimensions. It should also be noted that in a number of cases such models
 are exactly solvable at the classical level.

 One of the models of two-dimensional gravity which has been extensively
 discussed recently (at both the classical $^{13-16}$ and quantum$^{17-20}$
 levels) is the model$^{7}$ originally proposed in the context of bosonic
 string theory with dynamical torsion$^{21}$ in order to address certain
 difficulties of string theory. Thus, the study of Ref.~21 showed (using
 the path integral approach) that there is no critical dimension for the
 string with dynamical torsion. Note also that the model$^{7}$ presents the
 most general theory of two-dimensional $R^2$-gravity with independent
 dynamical torsion that leads to second-order equations of motion for the
 zweibein and Lorentz connection. The equations of motion of the model have
 been analyzed in Refs.~7, 13, 14 (in conformal$^{7,13}$ and
 light-cone$^{14}$ gauges) where their complete integrability was
 demonstrated.

 At the same time, the model$^{7}$ contains solutions with constant curvature
 and zero torsion, thus incorporating a number of other two-dimensional
 gravity models$^{1,2,9}$ (whose actions, however, do not admit, as compared
 to that of Ref.~7, of purely geometric interpretation).

 The Hamiltonian structure of gauge symmetries of the model$^{7}$ has been
 studied in Ref.~16 and its canonical quantization, in Refs.~17, 18.

 Our interest in the theory of two-dimensional gravity with dynamical
 torsion$^{7}$ is due to the recent paper$^{20}$ whose authors made an
 attempt at quantizing a model (with auxiliary fields) they suggested, which
 is classically equivalent to that of Ref.~7. To quantize this theory (in its
 formulation with the algebra of gauge transformations open of-shell) the
 authors$^{20}$ use a modification of the Faddeev--Popov rules; however, in
 our opinion, they failed to give a consistent procedure for constructing the
 quantum action. Notice that the classical equivalence of two theories
 generally does not imply their equivalence at the quantum level (see, for
 example, Ref.~[22]). In this connection, the present paper deals with the
 treatment of the original model$^{7}$ in different versions of Lagrangian
 quantization, namely in the BV$^{23}$ and BLT$^{24}$ quantization schemes
 for general gauge theories as well as in the background field method (for
 details see, for example, Ref.~25).

 We use De Witt's condensed notations$^{26}$ in general formulas of
 Refs.~23, 24. The Grassmann parity and ghost number of a certain quantity
 $A$ are denoted $\varepsilon(A)$, ${\rm gh}(A)$ respectively.

 For indices of quantities transforming by the Lorentz group, we use Latin
 characters: $i$, $j$, $k\ldots$ $(i=0,1)$; $\varepsilon^{ij}$ is a constant
 antisymmetric second-rank pseudo-tensor subject to the normalization
 condition $\varepsilon^{01}=1$. Greek characters stand for indices of
 quantities transforming as (pseudo-)tensors under the general coordinates
 transformations: $\lambda$, $\mu$, $\nu\ldots$ $(\lambda=0,1)$; given this
 $\epsilon^{\mu\nu}=-\epsilon^{\nu\mu}$ ($\epsilon^{01}=1$). Derivatives with
 respect to fields are understood as the right-hand, and those with respect
 to sources and antifields, as the left-hand ones.
\section{Two-dimensional Gravity with Dynamical Torsion}
 The theory$^{7}$ of two-dimensional gravity with dynamical torsion is
 described in terms of the zweibein and Lorentz connection
 $(e^i_\mu,\omega_\mu)$ by the action
\begin{equation}
 {\cal S}(e^i_\mu,\omega_\mu) = \int {d^2}x\;e\bigg(\frac{1}{16\alpha}
 R_{\mu\nu}{}^{ij}R^{\mu\nu}{}_{ij}-\frac{1}{8\beta}T_{\mu\nu}{}^{i}
 T^{\mu\nu}{}_{i}-\gamma\bigg),
\end{equation}
 where $\alpha$, $\beta$, $\gamma$ are constant parameters. In Eq.~(1), the
 Latin indices are lowered with the help of the Minkowski metric $\eta_{ij}$
 $(+,-)$, and the Greek indices, with the help of the metric tensor
 $g_{\mu\nu}=\eta_{ij}e^i_\mu e^j_\nu$. Besides, the following notations are
 used:
\begin{eqnarray*}
 e&=&{\rm det}e^i_\mu,\\
 R_{\mu\nu}{}^{ij}&=&\varepsilon^{ij}R_{\mu\nu},\;\;\;
 R_{\mu\nu}=\partial_\mu\omega_\nu-(\mu\leftrightarrow\nu),\\
 T_{\mu\nu}{}^{i}&=&
 \partial_\mu e^i_\nu+\varepsilon^{ij}\omega_\mu e_{\nu j}-
 (\mu\leftrightarrow\nu).
\end{eqnarray*}
 The action (1) is invariant under the local Lorentz transformations
 $e^i_\mu\rightarrow e^{'i}_\mu$, $\omega_\mu\rightarrow \omega^{'}{}_\mu$
\begin{eqnarray}
 e^{'i}_\mu&=&(\Lambda e_\mu)^i,\nonumber\\
 \\
 (\Omega^{'}{}_\mu)^i_j&=&(\Lambda\Omega_\mu\Lambda ^{-1})^i_j+
 (\Lambda\partial_\mu\Lambda ^{-1})^i_j,\;\;\;
 (\Omega_\mu)^i_j=\varepsilon^{ik}\eta_{kj}\omega_\mu,\nonumber
\end{eqnarray}
 or infinitesimally (with the parameter $\zeta$)
\begin{eqnarray}
 \delta e^i_\mu=\varepsilon^{ij}e_{\mu j}\zeta,\;\;\;
 \delta\omega_\mu=-\partial_\mu\zeta,
\end{eqnarray}
 as well as under the general coordinates transformations
\begin{eqnarray}
 x&\rightarrow&x^{'}=x^{'}(x),\nonumber\\
 e^i_\mu&\rightarrow&e^{'i}_\mu(x^{'})=\frac{\partial x^\lambda}
 {\partial x^{'\mu}}e^i_\lambda(x),\\
 \omega_\mu&\rightarrow& \omega^{'}_\mu(x^{'})=\frac{\partial
 x^\lambda}{\partial x^{'\mu}}\omega_\lambda(x),\nonumber
\end{eqnarray}
 and, consequently, under the corresponding infinitesimal transformations
 (with the parameters $\xi^\mu$)
\begin{eqnarray}
 \delta e^i_\mu&=&e^i_\nu\partial_\mu\xi^\nu+(\partial_\nu
 e^i_\mu)\xi^\nu,\nonumber\\
 \\
 \delta\omega_\mu&=&\omega_\nu\partial_\mu\xi^\nu+(\partial_\nu
 \omega_\mu)\xi^\nu.\nonumber
\end{eqnarray}
 Added together, the gauge transformations (3), (5) form a closed algebra
\begin{eqnarray}
 {[}\delta_{\zeta(1)},\;\delta_{\zeta(2)}{]}&=&0,\nonumber\\
 {[}\delta_{\xi(1)},\;\delta_{\xi(2)}{]}&=&\delta_{\xi(1,2)},\\
 {[}\delta_\zeta,\;\delta_\xi{]}&=&\delta_{\zeta^{'}},\nonumber
\end{eqnarray}
 where
\[
 \xi^\mu{}_{(1,2)}=\xi^\nu{}_{(1)}\partial_\nu\xi^\mu{}_{(2)}-
 (\partial_\nu\xi^\mu{}_{(1)})\xi^\nu{}_{(2)},\;\;\;
 \zeta^{'}=(\partial_\mu\zeta)\xi^\mu.\nonumber
\]
 The action (1) (when $e\neq 0$) admits of the representation
\[
 {\cal S}(e^i_\mu,\omega_\mu)=\int {d^2}x\;\bigg(\frac{1}{4e\alpha}F^2+
 \frac{1}{4e\beta}T_iT^i-e\gamma\bigg),
\]
 where
\[
 F=\frac{1}{2}\epsilon^{\mu\nu}R_{\mu\nu},\;\;\;
 T^i=\frac{1}{2}\epsilon^{\mu\nu}T_{\mu\nu}{}^i
\]
 and is equivalent$^{20}$ to the action $\tilde{{\cal S}}=
 \tilde{{\cal S}}(e^i_\mu,\omega_\mu,\varphi,\varphi_i)$
\begin{eqnarray}
 \tilde{{\cal S}}(e^i_\mu,\omega_\mu,\varphi,\varphi_i)=
 \int{d^2}x\;\{(\varphi F+\varphi_iT^i)-e(\alpha\varphi^2+
 \beta\varphi_i\varphi^i+\gamma)\}
\end{eqnarray}
 after the auxiliary fields $\varphi$, $\varphi_i$ have been eliminated
 with the help of the equations of motion of $\tilde{{\cal S}}$
\begin{eqnarray*}
 \frac{\delta\tilde{{\cal S}}}{\delta e^i_\mu}&=&
 \epsilon^{\mu\nu}\{\partial_\nu\varphi_i+\varepsilon_{ij}\omega_\nu
 \varphi^j+\varepsilon_{ij}e^j_\nu(\alpha\varphi^2+\beta\varphi_k
 \varphi^k+\gamma)\},\\
 \frac{\delta\tilde{{\cal S}}}{\delta\omega_\mu}&=&
 \epsilon^{\mu\nu}(\partial_\nu\varphi+\varepsilon^{ij}\varphi_{i}
 e_{\nu j}),\\
 \frac{\delta\tilde{{\cal S}}}{\delta\varphi}&=&F-2\alpha e\varphi,\\
 \frac{\delta\tilde{{\cal S}}}{\delta\varphi_i}&=&T^i-2\beta e\varphi^i.
\end{eqnarray*}
 The action (7) is invariant under the complex of the local Lorentz
 transformations (2) of the initial fields $e^i_\mu$, $\omega_\mu$
 and the transformations of the auxiliary fields $\varphi$, $\varphi_i$
\begin{equation}
 \varphi^{'}=\varphi,\;\;\;\varphi^{'}_i=(\Lambda^{-1})^j_i\varphi_j,
\end{equation}
 as well as under the complex of the general coordinates transformations
 (4) and transformations of the form
\begin{equation}
 \varphi^{'}(x^{'})=\varphi(x),\;\;\;
 \varphi^{'}_i(x^{'})=\varphi_i(x).
\end{equation}
 Infinitesimally, (2), (8) and (4), (9) imply the gauge transformations
 (with the parameters $\zeta$, $\xi^\mu$)
\begin{equation}
\begin{array}{rcl}
 \delta e^i_\mu&=&\varepsilon^{ij}e_{\mu j}\zeta+
 e^i_\nu\partial_\mu\xi^\nu+(\partial_\nu e^i_\mu)\xi^\nu,\\
 \delta\omega_\mu&=&-\partial_\mu\zeta+
 \omega_\nu\partial_\mu\xi^\nu+(\partial_\nu
 \omega_\mu)\xi^\nu,\\
 \delta\varphi&=&(\partial_\mu\varphi)\xi^\mu,\\
 \delta\varphi_i&=&\varepsilon_{ij}\varphi^j\zeta+
 (\partial_\nu\varphi_i)\xi^\nu,
\end{array}
\end{equation}
 which form a closed algebra of the form (6).

 Notice that the action (7) is also invariant under the infinitesimal
 transformations (with the parameters $\bar{\zeta}$, $\bar{\xi}^i$)
\begin{equation}
\begin{array}{rcl}
 \bar{\delta}e^i_\mu&=&\varepsilon^{ij}e_{\mu j}\bar{\zeta}+
 \partial_\mu\bar{\xi}^i+\varepsilon^{ij}\omega_\mu\bar{\xi}_j+
 2\beta\varepsilon_{kl}\varphi^ie^k_\mu\bar{\xi}^l,\\
 \bar{\delta}\omega_\mu&=&-\partial_\mu\bar{\zeta}+
 2\alpha\varepsilon_{kl}\varphi e^k_\mu\bar{\xi}^l,\\
 \bar{\delta}\varphi&=&-\varepsilon^{ij}\varphi_i\bar{\xi}_j,\\
 \bar{\delta}\varphi_i&=&\varepsilon_{ij}\varphi^j\bar{\zeta}-
 \varepsilon_{ij}(\alpha\varphi^2+\beta\varphi_k\varphi^k+\gamma)
 \bar{\xi}^j,
\end{array}
\end{equation}
 whose algebra is open on the extremals of $\tilde{{\cal S}}$ (7)
\begin{eqnarray*}
 {[}\bar{\delta}_{\bar{\zeta}(1)},\;\bar{\delta}_{\bar{\zeta}(2)}{]}&=&0,
 \nonumber\\
 {[}\bar{\delta}_{\bar{\zeta}},\;\bar{\delta}_{\bar{\xi}}{]}&=&
 \bar{\delta}_{\bar{\xi}^{'}},\\
 {[}\bar{\delta}_{\bar{\xi}(1)},\;\bar{\delta}_{\bar{\xi}(2)}{]}&=&
 \bar{\delta}_{\bar{\zeta}(1,2)}+\bar{\delta}_{\bar{\xi}(1,2)}+
 2\epsilon_{\mu\nu}\varepsilon_{kl}\bar{\xi}^k{}_{(1)}
 \bar{\xi}^l{}_{(2)}
 \bigg(\alpha\frac{\delta\tilde{{\cal S}}}{\delta\omega_\mu}
 \frac{\delta}{\delta\omega_\nu}+\beta\eta^{ij}
 \frac{\delta\tilde{{\cal S}}}{\delta e^i_\mu}
 \frac{\delta}{\delta e^j_\nu}\bigg),
\end{eqnarray*}
 where
\begin{eqnarray*}
 \bar{\xi}^{'i}&=&-\varepsilon^{ij}\bar{\zeta}\bar{\xi}_j,\\
 \bar{\zeta}_{(1,2)}&=&-2\alpha\varepsilon_{ij}\bar{\xi}^i{}_{(1)}
 \bar{\xi}^j{}_{(2)}\varphi,\\
 \bar{\xi}^i{}_{(1,2)}&=&2\beta\varepsilon_{jk}\bar{\xi}^j{}_{(1)}
 \bar{\xi}^k{}_{(2)}\varphi^i.
\end{eqnarray*}
 The transformations (11) correspond to a set of generators which is
 equivalent to that of Eq.~(10) and coincide, for the choice of the
 parameters $\bar{\zeta}=\zeta-\omega_\mu\xi^\mu$,
 $\bar{\xi}^i=e^i_\mu\xi^\mu$, with (10) on-shell
\begin{eqnarray*}
 \bar{\delta}e^i_\mu&=&{\delta}e^i_\mu-\epsilon_{\mu\nu}
 \frac{\delta\tilde{{\cal S}}}{\delta\varphi_i}\xi^\nu,\\
 \bar{\delta}\omega_\mu&=&{\delta}\omega_\mu-\epsilon_{\mu\nu}
 \frac{\delta\tilde{{\cal S}}}{\delta\varphi}\xi^\nu,\\
 \bar{\delta}\varphi&=&{\delta}\varphi+\epsilon_{\mu\nu}
 \frac{\delta\tilde{{\cal S}}}{\delta\omega_\mu}\xi^\nu,\\
 \bar{\delta}\varphi_i&=&{\delta}\varphi_i+\epsilon_{\mu\nu}
 \frac{\delta\tilde{{\cal S}}}{\delta e^i_\mu}\xi^\nu.
\end{eqnarray*}

 Notice that the result of Lagrangian quantization generally depends on the
 way the initial gauge theory is chosen from the class of equivalent theories
 and, in particular, on the choice of generators of gauge transformations.
 Thus, the study of Ref.~22 has demonstrated that a theory with a unitary
 $S$-matrix for a certain choice of generators may prove non-unitary for
 another choice. In this connection, we further concentrate on the original
 model (1), (3), (5), (6).

\section{Quantization of the Model}
 Let us consider the original model (1), (3), (5), (6) in the framework of
 the BV quantization formalism$^{23}$ for irreducible gauge theories.
 To this end, we first introduce the complete configuration space $\phi^A$.
 It is constructed by extending the initial space of the fields
 $(e^i_\mu,\omega_\mu)$ with the help of the Faddeev--Popov ghosts
 ($\overline{C}$, $C$, $\overline{C}^\mu$, $C^\mu$) and the Lagrangian
 multipliers ($b$, $b^\mu$) according to the number of the gauge parameters
 in Eqs.~(3), (5) (for $\zeta$, $\xi^\mu$ respectively). The Grassmann
 parity and ghost numbers of the fields $\phi^A$
\[
 \phi^A=(e^i_\mu,\;\omega_\mu;\;b,\;b^\mu;\;\overline{C},\;C,\;
 \overline{C}^\mu,\;C^\mu)
\]
 have the form
\begin{eqnarray*}
 \varepsilon(e^i_\mu)&=&\varepsilon(\omega_\mu)=\varepsilon(b)=
 \varepsilon(b^\mu)=0,\\
 \varepsilon(\overline{C})&=&\varepsilon(C)=
 \varepsilon(\overline{C}^\mu)=\varepsilon(C^\mu)=1,\\
 {\rm gh}(e^i_\mu)&=&{\rm gh}(\omega_\mu)={\rm gh}(b)=
 {\rm gh}(b^\mu)=0,\\
 {\rm gh}(\overline{C})&=&{\rm gh}(\overline{C}^\mu)=-1,\;\;\;
 {\rm gh}(C)={\rm gh}(C^\mu)=1.
\end{eqnarray*}
 Besides, we introduce for the fields ${\phi}^A$ the set of the antifields
 ${\phi}_{A}^{*}$
\[
 \phi_{A}^*=(e^{*\mu}_i,\;\omega^{*\mu};\;b^*,\;b^*_\mu;\;
 \overline{C}^*,\;C^*,\;\overline{C}^*_\mu,\;C^*_\mu)
\]
 with the following distribution of the Grassmann parity and ghost number:
\[
 \varepsilon(\phi_{A}^*)=\varepsilon(\phi^A)+1,\;\;\;
 {\rm gh}(\phi_{A}^*)=-1-{\rm gh}(\phi^A).
\]
\hspace*{\parindent}
 As is well-known, the generating functional $Z(J)$ of Green's functions for
 the fields $\phi^A$ can be represented, within the BV quantization, in the
 form of the following functional integral:
\begin{eqnarray}
 Z(J)=\int d\phi\;d\phi^*\;d\lambda\exp\bigg\{\frac{i}{\hbar}\bigg[S
 (\phi,\phi^*)+\bigg(\phi^*_A-\frac{\delta\Psi}
 {\delta\phi^A}\bigg)\lambda^A+J_A\phi^A\bigg]\bigg\}.
\end{eqnarray}
 In Eq.~(12), $J_A$ are the sources to the fields $\phi^A$
\[
 \varepsilon(J_A)=\varepsilon(\phi^A),\;\;\;
 {\rm gh}(J_A)=-{\rm gh}(\phi^A),
\]
 $\lambda^A$ are auxiliary fields
\[
 \varepsilon(\lambda^A)=\varepsilon(\phi^*_A),\;\;\;
 {\rm gh}(\lambda^A)=-{\rm gh}(\phi^*_A),
\]
 $\Psi=\Psi(\phi)$ is the gauge fermion, and $S=S(\phi,\phi^*)$ is a boson
 functional satisfying the generating equation
\begin{equation}
 \frac{\delta S}{\delta\phi^A}\frac{\delta S}{\delta\phi^*_A}
 =i\hbar\frac{\delta^2S}{\delta\phi^A\delta\phi^*_A}
\end{equation}
 with the boundary condition
\begin{equation}
 \left.S\right|_{\phi^*=\hbar=0}={\cal S},
\end{equation}
 where ${\cal S}$ is the initial classical action.

 The solution of the generating equation (13) satisfying the boundary
 condition (14) for the model (1), (3), (5), (6) can be represented as
 a functional $S=S(\phi,\phi^*)$ linear in the antifields (we assume a
 regularization of dimensional type)
\begin{eqnarray}
 S&=&S_{\rm min}+\int d^2x\;(\overline{C}^*b+
 \overline{C}^*_\mu b^\mu),\nonumber\\
 S_{\rm min}&=&{\cal S}+
 \int d^2x\;\{e^{*\mu}_iX_{1\mu}^i+\omega^{*\mu}X_{1\mu}+C^*X_2+
 C^*_\mu X^\mu_2\},
\end{eqnarray}
where
\begin{eqnarray*}
 X_{1\mu}^i&=&\varepsilon^{ij}e_{\mu j}C+
 (\partial_\lambda e^i_\mu)C^\lambda+e^i_\lambda\partial_\mu C^
 \lambda,\\
 X_{1\mu}&=&-\partial_\mu C+(\partial_\lambda\omega_\mu)C^\lambda+
 \omega_\lambda\partial_\mu C^\lambda,\\
 X_2&=&\frac{1}{2}(\partial_\mu C)C^\mu,\;\;\;
 X^\mu_2=C^\lambda\partial_\lambda C^\mu\
\end{eqnarray*}
 We choose the gauge fermion $\Psi=\Psi(\phi)$ in the form
\begin{eqnarray}
 \Psi&=&\int d^2x\;(\overline{C}\chi+
 \overline{C}^\mu\chi_\mu),\nonumber\\
 \\
 \chi&=&\eta^{\mu\nu}\partial_\mu\omega_\nu,\;\;\;
 \chi_\mu=\eta^{\lambda\nu}e_{\lambda i}\partial_\mu e^i_\nu,
 \nonumber
\end{eqnarray}
 where $\eta^{\mu\nu}$ $(+,-)$ is the metric of the two-dimensional
 Minkowski space.

 By virtue of Eqs.~(15), (16), integrating over the variables $\lambda^A$,
 $\phi^*_A$ in Eq.~(12) yields the following representation of the generating
 functional of Green's functions:
\begin{eqnarray}
 Z(J)=\int d\phi\;\exp\bigg\{\frac{i}{\hbar}\bigg({\cal S}+S_{\rm
 GF}(\phi)+ S_{\rm GH}(\phi)+J_A\phi^A\bigg)\bigg\},
\end{eqnarray}
 where
\begin{eqnarray*}
 S_{\rm GF}&=&\int d^2x\;(\chi b+\chi_\mu b^\mu),\\
 S_{\rm GH}&=&\int d^2x\;\bigg\{\eta^{\mu\nu}\partial_\nu\overline{C}
 \bigg(\partial_\mu C-(\partial_\lambda\omega_\mu)C^\lambda-
 \omega_\lambda\partial_\mu C^\lambda\bigg)-\\&&
 -\bigg(\eta^{\mu\nu}\overline{C}^\lambda\partial_\lambda e^i_\nu-
 \eta^{\lambda\mu}\partial_\nu(\overline{C}^\nu e^i_\lambda)\bigg)
 \bigg(\varepsilon_{ij}e^j_\mu C
 -(\partial_\sigma e_{\mu i})C^\sigma
 -e_{\sigma i}\partial_\mu C^\sigma\bigg)\bigg\}.
\end{eqnarray*}

 Consider now the model (1), (3), (5), (6) in the Lagrangian version of
 the BLT quantization$^{24}$ of gauge theories. Notice that the
 Faddeev--Popov ghosts are then combined into the $Sp$(2)-doublets ($a=1,2$)
\[
 C^a=(\overline{C},C),\;\;\;C^{\mu a}=(\overline{C}^\mu,C^\mu),
\]
 and the fields $\phi^A$ of the complete configuration space are
 supplemented by the sets of the antifields $\phi^*_{Aa}$, $\bar{\phi}_A$
\begin{eqnarray*}
 \phi^*_{Aa}&=&(e^{*\mu}_{ia},\;\omega^{*\mu}_a;\;b^*_a,\;b^*_{\mu a};\;
 C^*_{ba},\;C^*_{\mu ba}),\\
 \bar{\phi}_A&=&(\overline{e}^\mu_i,\;\overline{\omega}^\mu;\;
 \overline{b},\;\overline{b}_\mu;\;\overline{C}_a,\;\overline{C}_{\mu a})
\end{eqnarray*}
 with
\begin{eqnarray*}
 \varepsilon(\phi^*_{Aa})&=&\varepsilon(\phi^A)+1,\;\;\;
 {\rm gh}(\phi^*_{Aa})=(-1)^a-{\rm gh}(\phi^A),\\
 \varepsilon(\bar{\phi}_A)&=&\varepsilon(\phi^A),\;\;\;
 {\rm gh}(\bar{\phi}_A)=-{\rm gh}(\phi^A).
\end{eqnarray*}
 The generating functional $Z(J)$ of Green's functions within the Lagrangian
 version of the BLT method can be represented in the form$^{24}$
\begin{eqnarray}
 Z(J)&=&\int d\phi\;d\phi^{*}_a\;d\bar{\phi}\;d\pi^a\;d\lambda\exp
 \bigg\{\frac{i}{\hbar}\bigg[S(\phi,\phi^{*},\bar{\phi})+\phi^{*}_{Aa}
 \pi^{Aa}+\nonumber\\&&
 +\bigg(\bar{\phi}_A-\frac{\delta F}
 {\delta\phi^A}\bigg)\lambda^A-\frac{1}{2}\varepsilon_{ab}\pi^{Aa}
 \frac{\delta^2F}{\delta\phi^A\delta\phi^B}\pi^{Bb}+
 J_A\phi^A\bigg]\bigg\}\;.
 \label{gfGf}
\end{eqnarray}
 In Eq.~(18), $\varepsilon_{ab}$ is an antisymmetric tensor
 ($\varepsilon_{12}=-1$); $\pi^{Aa}$, $\lambda^A$ are auxiliary fields
\begin{eqnarray*}
 \varepsilon(\pi^{Aa})&=&\varepsilon_A+1,\;\;
 {\rm gh}(\pi^{Aa})=-(-1)^a+{\rm gh}(\phi^A),\\
 \varepsilon(\lambda^A)&=&\varepsilon_A,\;\;{\rm gh}(\lambda^A)=
 {\rm gh}(\phi^A),
\end{eqnarray*}
 $F=F(\phi)$ is the gauge fixing boson functional, and
 $S=S(\phi,\phi^*,\overline{\phi})$ is a boson functional
 satisfying the generating equations of the BLT formalism$^{24}$
\begin{equation}
 \frac{\delta S}{\delta\phi^A}\frac{\delta S}{\delta\phi_{Aa}^\ast}
 + \varepsilon^{ab}\phi_{Ab}^\ast \frac{\delta S}{\delta\overline{\phi}_A}=
 i\hbar\frac{\delta^2S}{\delta\phi^A\delta\phi_{Aa}^\ast}
 \label{eq}
\end{equation}
 with the boundary condition
\begin{equation}
 \left.S\right|_{\phi^\ast=\overline{\phi}=\hbar=0}={\cal S}.
 \label{bc}
\end{equation}

 The solution of the generating equations (19) with the boundary
 condition (20) for the theory in question can be chosen as a
 functional linear in the antifields $\phi^*_{Aa}$, $\bar{\phi}_A$
\begin{eqnarray}
 S={\cal S}&+&\int d^2x\bigg\{
 e^{*\mu}_{ia}X^{ia}_{1\mu}+\omega^{*\mu}_aX^{a}_{1\mu}
 +b^*_aX^{a}_{2}+b^*_{\mu a}X^{\mu a}_{2}\nonumber\\
 &+&C^*_{ba}X^{ab}_{3}+C^*_{\mu ba}X^{\mu ab}_{3}
 +\overline{e}^\mu_iY^i_{1\mu}+\overline{\omega}^\mu Y_{1\mu}
 +\overline{C}_aY^a_2+\overline{C}_{\mu a}Y^{\mu a}_2
 \bigg\},
 \label{sol}
\end{eqnarray}
 where
\begin{eqnarray*}
 X^{ia}_{1\mu}&=&\varepsilon^{ij}e_{\mu j}C^a+
 (\partial_\lambda e^i_\mu)C^{\lambda a}+
 e^i_{\lambda}\partial_\mu C^{\lambda a},\\
 X^{a}_{1\mu}&=&-\partial_\mu C^a+(\partial_\lambda\omega_\mu)C^{\lambda a}
 +\omega_\lambda\partial_\mu C^{\lambda a},\\
 X^{a}_{2}&=&\frac{1}{2}
 \bigg(
 C^{\mu a}\partial_\mu b
 -\frac{1}{6}\varepsilon_{bc}C^{\mu c}\partial_\mu(C^{\lambda a}
 \partial_\lambda C^b)
 \bigg),\\
 X^{\mu a}_{2}&=&\frac{1}{2}
 \bigg(
 (b^\lambda\partial_\lambda C^{\mu a}-C^{\lambda a}\partial_\lambda b^\mu)
 +\frac{1}{6}\varepsilon_{bc}
 [(C^{\sigma b}\partial_\sigma C^{\lambda a}+C^{\sigma a}\partial_\sigma
 C^{\lambda b})\partial_\lambda C^{\mu c}\\
 &&-C^{\lambda c}\partial_\lambda(C^{\sigma b}\partial_\sigma C^{\mu a}
 +C^{\sigma a}\partial_\sigma C^{\mu b})]
 \bigg),\\
 X^{\mu ab}_{3}&=&-\varepsilon^{ab}b+
 \frac{1}{2}(\partial_\mu C^b)C^{\mu a},\\
 X^{\mu ab}_{3}&=&-\varepsilon^{ab}b^\mu+
 \frac{1}{2}(C^{\lambda b}\partial_\lambda C^{\mu a}+
 C^{\lambda a}\partial_\lambda C^{\mu b}),\\
 Y^i_{1\mu}&=&\varepsilon^{ij}be_{\mu j}+b^\lambda\partial_\lambda e^i_\mu
 +e^i_\lambda\partial_\mu b^\lambda
 +\frac{1}{2}\varepsilon_{ab}
 \bigg(
 (e^i_\mu C^b+\varepsilon^{ij}C^{\lambda b}\partial_\lambda e_{\mu j}
 +\varepsilon^{ij}e_{\lambda j}\partial_\mu C^{\lambda b})C^a\\
 &&-C^{\lambda a}\partial_\lambda(\varepsilon^{ij}e_{\mu j}C^b
 +C^{\sigma b}\partial_\sigma e^i_\mu+e^i_\sigma\partial_\mu C^{\sigma b})\\
 &&+(\varepsilon^{ij}e_{\lambda j}C^b
 +(\partial_\sigma e^i_\lambda)C^{\sigma b}
 +e^i_\sigma\partial_\lambda C^{\sigma b})\partial_\mu C^{\lambda a}\bigg),\\
 Y_{1\mu}&=&-\partial_\mu b+b^\lambda\partial_\lambda\omega_\mu
 +\omega_\lambda\partial_\mu b^\lambda
 -\frac{1}{2}\varepsilon_{ab}
 [
 C^{\lambda a}\partial_\lambda
 (C^{\sigma b}\partial_\sigma\omega_\mu+\omega_\sigma\partial_\mu
 C^{\sigma b}-\partial_\mu C^b)\\
 &&-(C^{\sigma b}\partial_\sigma\omega_\lambda
 +\omega_\sigma\partial_\lambda C^{\sigma b}-\partial_\lambda C^b)
 \partial_\mu C^{\lambda a}
 ],\\
 Y^a_2&=&-2X^a_2,\\
 Y^{\mu a}_2&=&-2X^{\mu a}_2.
\end{eqnarray*}
 We choose the gauge boson $F=F(\phi)$ in the form ($p$, $q$ are constant
 parameters)
\[
 F=\int d^2x\bigg\{\frac{p}{2}\eta_{ij}\eta^{\mu\nu}e^i_\mu e^j_\nu
 +\frac{q}{2}\eta^{\mu\nu}\omega_\mu\omega_\nu\bigg\}.
\]
 Then, integrating in Eq.~(18) over the variables $\lambda^A$, $\pi^{Aa}$,
 $\overline{\phi}_A$, $\phi^*_{Aa}$ and taking Eq.~(21) into account, we have
 the following representation for $Z(J)$:
\begin{eqnarray}
 Z(J)=\int d\phi\exp\bigg\{\frac{i}{\hbar}\bigg(
 {\cal S}+S_{\rm GF}(\phi)+S_{\rm GH}(\phi)+J_A\phi^A\bigg)\bigg\},
\end{eqnarray}
where
\begin{eqnarray*}
 S_{\rm GF}&=&\int d^2x(\chi b+\chi_\mu b^\mu),\\
 S_{\rm GH}&=&\int d^2x
 \bigg\{
 p\bigg(
 \eta^{\mu\nu}(e^i_\lambda\partial_\nu C^{\lambda a}
 -e^i_\nu\partial_\lambda C^{\lambda a})\\
 &&+\eta^{\lambda\nu}e^i_\nu\partial_\lambda C^{\mu a}
 \bigg)
 \bigg(
 \varepsilon_{ij}C^be^j_\mu-\eta_{ij}(C^{\sigma b}\partial_\sigma e^j_\mu
 +e^j_\sigma\partial_\mu C^{\sigma b})
 \bigg)\\
 &&-q
 \bigg(
 \eta^{\mu\nu}(\partial_\nu C^a
 -\omega_\lambda\partial_\nu C^{\lambda a}
 +\omega_\nu\partial_\lambda C^{\lambda a})\\
 &&-\eta^{\lambda\nu}\omega_\nu\partial_\lambda C^{\mu a}
 \bigg)
 (\partial_\mu C^b-C^{\sigma b}\partial_\sigma\omega_\mu
 -\omega_\sigma\partial_\mu C^{\sigma b})
 \bigg\}
\end{eqnarray*}
 and
\begin{eqnarray*}
 \chi&=&q\eta^{\mu\nu}\partial_\mu\omega_\nu,\\
 \chi_\mu&=&\eta^{\lambda\nu}
 \bigg\{
 p
 \bigg(
 e_{\lambda i}\partial_\mu e^i_\nu-\partial_\lambda(e_{\mu i}e^i_\nu)
 \bigg)
 +q\bigg(
 \omega_\lambda\partial_\mu\omega_\nu-
 \partial_\lambda(\omega_\mu\omega_\nu)
 \bigg)
 \bigg\},\\
\end{eqnarray*}

 We finally consider the initial theory (1), (3), (5), (6) in the
 framework of the background field method (see, for example, Ref.~25).
 For this purpose, we first assign to the initial fields the set
 $(A,Q)$ of the background $A=(e^i_\mu,\omega_\mu)$ and quantum
 $Q=(q^i_\mu,q_\mu)$ fields. Secondly, we associate the initial gauge
 transformations (3), (5) with two kinds of transformations, namely
 background $\delta_B$ and quantum $\delta_Q$, by the rule
\begin{eqnarray}
\begin{array}{rcl}
 \delta_B e^i_\mu&=&\varepsilon^{ij}e_{\mu j}\zeta+
 e^i_\nu\partial_\mu\xi^\nu+(\partial_\nu e^i_\mu)\xi^\nu,\\
 \delta_B\omega_\mu&=&-\partial_\mu\zeta+
 \omega_\nu\partial_\mu\xi^\nu+(\partial_\nu
 \omega_\mu)\xi^\nu,\\
 \delta_B q^i_\mu&=&\varepsilon^{ij}q_{\mu j}\zeta+
 q^i_\nu\partial_\mu\xi^\nu+(\partial_\nu q^a_\mu)\xi^\nu,\\
 \delta_Bq_\mu&=&q_\nu\partial_\mu\xi^\nu+(\partial_\nu
 q_\mu)\xi^\nu,
\end{array}
\end{eqnarray}
\begin{eqnarray}
\begin{array}{rcl}
 \delta_Q e^i_\mu&=&\delta_Q\omega_\mu=0,\\
 \delta_Q q^i_\mu&=&\varepsilon^{ij}(e_{\mu j}+q_{\mu j})\zeta+
 (e^i_\nu+q^i_\nu)\partial_\mu
 \xi^\nu+(\partial_\nu e^i_\mu+\partial_\nu q^i_\mu)
 \xi^\nu,\\
 \delta_Q q_\mu&=&-\partial_\mu\zeta+
 (\omega_\nu+q_\nu)\partial_\mu\xi^\nu+
 (\partial_\nu\omega_\mu+\partial_\nu q_\mu)\xi^\nu.
\end{array}
\end{eqnarray}
 Clearly, the action ${\cal S}(A+Q)$ in Eq.~(1) is invariant under both kinds
 of transformations (23), (24).

 Following the background field method, we further introduce the analog
 $Z(J,A)$ of the generating functional of Green's functions (its relation to
 the standard generating functional has been established in Ref.~27)
\begin{eqnarray}
 Z(J,A)=\int
 dQ\;d\overline{C}\;dC\;\exp\bigg\{\frac{i}{\hbar}\bigg({\cal S}(A+Q)
 +S_{\rm GF}(A,Q)+S_{\rm GH}(A,Q;\overline{C},C)+JQ\bigg)\bigg\},
\end{eqnarray}
 where $J=(J^\mu_i,J^\mu)$ are the sources to the quantum fields
 $Q=(q^i_\mu,q_\mu)$. In Eq.~(25), $S_{\rm GH}=S_{\rm GH}(A,Q)$ is a
 functional constructed with the help of the gauge fixing functions
 $\chi$, $\chi_\mu$ (respectively for the parameters $\zeta$, $\xi^\mu$)
 according to the requirement of invariance under the background
 transformations, i.e. $\delta_B S_{\rm GH}=0$. Given this, the functional
 $S_{\rm GH}=S_{\rm GH}(A,Q;\overline{C},C)$ is constructed by the rule
\begin{eqnarray}
 S_{\rm GH}=\int d^2x\;(\overline{C}\delta_Q\chi+
 \overline{C}^\mu\delta_Q\chi_\mu),
\end{eqnarray}
 where in the transformations $\delta_Q$ we make the replacement
 $(\zeta,\xi^\mu)\rightarrow(C,C^\mu)$.

 Let us introduce the gauge-fixing conditions $\chi=\chi(A,Q)$,
 $\chi_\mu=\chi_\mu(A,Q)$ linear in the quantum fields $Q=(q^i_\mu,q_\mu)$
\begin{eqnarray}
 \chi&=&eg^{\mu\nu}\nabla_\mu q_\nu,\nonumber\\
 \\
 \chi_\mu&=&eg^{\lambda\nu}e_{\mu i}\nabla_\lambda q^i_\nu,
 \nonumber
\end{eqnarray}
 where $e$, $g^{\mu\nu}$ are constructed from the background fields
 $e^i_\mu$ ($g^{\mu\lambda}g_{\lambda\nu}=\delta^\mu_\nu$,
 $g_{\mu\nu}=\eta_{ij}e^i_\mu e^j_\nu$, $e={\rm det}e^i_\mu$),
 and the action of the covariant derivative $\nabla_\mu$ on an arbitrary
 (psedo-)tensor field
 $T^{\nu_1\ldots\nu_l}_{\mu_1\ldots\mu_k}
 {}^{j_1\ldots j_n}_{i_1\ldots i_m}$ is given by the rule
 ($(\Omega_\mu)^i_j=\varepsilon^{ik}\eta_{kj}\omega_\mu$)
\begin{eqnarray}
 \nabla_\mu
 T^{\nu_1\ldots\nu_l}_{\mu_1\ldots\mu_k}{}^{j_1\ldots j_n}_{i_1\ldots i_m}
 &=&\partial_\mu
 T^{\nu_1\ldots\nu_l}_{\mu_1\ldots\mu_k}{}^{j_1\ldots j_n}_{i_1\ldots i_m}
 -\Gamma^{\lambda}_{\mu\mu_1}
 T^{\nu_1\ldots\nu_l}_{\lambda\ldots\mu_k}{}^{j_1\ldots j_n}_
 {i_1\ldots i_m}
 -\ldots
 -\Gamma^{\lambda}_{\mu\mu_k}
 T^{\nu_1\ldots\nu_l}_{\mu_1\ldots\lambda}{}^{j_1\ldots j_n}_
 {i_1\ldots i_m}
 +\nonumber\\
 \nonumber\\&&
 +\Gamma^{\nu_1}_{\mu\lambda}
 T^{\lambda\ldots\nu_l}_{\mu_1\ldots\mu_k}{}^{j_1\ldots j_n}_
 {i_1\ldots i_m}
 +\ldots
 +
 \Gamma^{\nu_n}_{\mu\lambda}
 T^{\nu_1\ldots\lambda}_{\mu_1\ldots\mu_k}{}^{j_1\ldots j_n}_
 {i_1\ldots i_m}+\nonumber\\
 \nonumber\\&&
 +
 (\Omega_\mu)^{j_1}_p
 T^{\nu_1\ldots\nu_l}_{\mu_1\ldots\mu_k}{}^{p\ldots j_n}_{i_1\ldots i_m}
 +\ldots
 +
 (\Omega_\mu)^{j_n}_p
 T^{\nu_1\ldots\nu_l}_{\mu_1\ldots\mu_k}{}^{j_1\ldots p}_{i_1\ldots
 i_m}-\nonumber\\
 \nonumber\\&&
 -
 (\Omega_\mu)^{p}_{i_1}
 T^{\nu_1\ldots\nu_l}_{\mu_1\ldots\mu_k}{}^{j_1\ldots j_n}_{p\ldots i_m}
 -\ldots
 -
 (\Omega_\mu)^{p}_{i_m}T^{\nu_1\ldots\nu_l}_{\mu_1\ldots\mu_k}{}^
 {j_1\ldots j_n}_{i_1\ldots p}\;.
\end{eqnarray}
 In Eq.~(28), $\Gamma^{\lambda}_{\mu\nu}$ are constructed with the help of
 the metric $g_{\mu\nu}$
\[
 \Gamma^{\lambda}_{\mu\nu}=\frac{1}{2}g^{\lambda\sigma}(
 \partial_\nu g_{\mu\sigma}+\partial_\mu g_{\nu\sigma}-
 \partial_\sigma g_{\mu\nu}).
\]
 Clearly, the covariant derivative $\nabla_\mu$ (28) satisfies the property
\[
 \nabla_\mu(FG)=F\nabla_\mu G+(\nabla_\mu F)G,
\]
 where $F$, $G$ are arbitrary (psedo-)tensor fields. At the same time,
 one readily establishes the relations
\[
 \nabla_\sigma g_{\mu\nu}=\nabla_\sigma g^{\mu\nu}=0.
\]
 We choose for $S_{\rm GF}$ the functional ($p$, $q$ are some numbers)
\begin{equation}
 S_{\rm GF}=\frac{1}{2}\int
 d^2x\;e(p\chi^2+q\chi_\mu\chi^\mu),
\end{equation}
 invariant, by construction, under the local Lorentz transformations of the
 form
\begin{eqnarray}
 e^{'i}_\mu&=&(\Lambda e_\mu)^i,\;\;\;
 q^{'i}_\mu=(\Lambda q_\mu)^i,
 \nonumber\\
 \\
 (\Omega^{'}{}_\mu)^i_b&=&(\Lambda\Omega_\mu\Lambda ^{-1})^i_j+
 (\Lambda\partial_\mu\Lambda ^{-1})^i_j,\;\;\;
 q^{'}_\mu=q_\mu,\nonumber
\end{eqnarray}
 as well as under the general coordinates transformations
\begin{eqnarray}
 e^{'i}_\mu(x^{'})&=&\frac{\partial x^\lambda}
 {\partial x^{'\mu}}e^i_\lambda(x),\;\;\;
 \omega^{'}_\mu(x^{'})=\frac{\partial
 x^\lambda}{\partial x^{'\mu}}\omega_\lambda(x),\nonumber\\
 \\
 q^{'i}_\mu(x^{'})&=&\frac{\partial x^\lambda}
 {\partial x^{'\mu}}q^i_\lambda(x),\;\;\;
 q^{'}_\mu(x^{'})=\frac{\partial
 x^\lambda}{\partial x^{'\mu}}q_\lambda(x).\nonumber
\end{eqnarray}
 At the same time, the infinitesimal form of the transformations
 (30) and (31) coincides with the background transformations (23).

 Notice that the (non-vanishing) quantum transformations (24) can be
 represented, with allowance for the definition (28), in the form
 ($\zeta\rightarrow C$, $\xi^\mu\rightarrow C^\mu$)
\begin{eqnarray*}
 \delta_Q q^i_\mu&=&\varepsilon^{ij}(e_{\mu j}+q_{\mu j})C+
 (e^i_\nu+q^i_\nu)\nabla_\mu C^\nu+
 (\nabla_\nu e^i_\mu+\nabla_\nu q^i_\mu)C^\nu
 -\varepsilon^{ij}\omega_\nu(e_{\mu j}+q_{\mu b})C^\nu,\\
 \\
 \delta_Q q_\mu&=&-\nabla_\mu C+
 (\omega_\nu+q_\nu)\nabla_\mu C^\nu+
 (\nabla_\nu\omega_\mu+\nabla_\nu q_\mu)C^\nu.
\end{eqnarray*}
 Then the functional $S_{\rm GH}$ (26) takes on the form
\begin{eqnarray}
 S_{\rm GH}&=&\int d^2x\;e\{
 -\overline{C}\nabla_\mu\nabla^\mu C+
 \overline{C}\nabla^\mu[(\nabla_\nu\omega_\mu+
 \nabla_\nu q_\mu)C^\nu
 +(\omega_\nu+q_\nu)\nabla_\mu C^\nu]+\nonumber\\&&
 \nonumber\\&&
 +\varepsilon^{ij}\overline{C}^\mu e_{\mu i}\nabla^\nu
 [(e_{\nu j}+q_{\nu j})(C-\omega_\lambda C^\lambda)]+\nonumber\\&&
 \nonumber\\&&
 +\overline{C}^\mu e_{\mu i}\nabla^\nu
 [(\nabla_\lambda e^i_{\nu}+\nabla_\lambda q^i_{\nu})C^\lambda+
 (e^i_{\lambda}+q^i_\lambda)\nabla_\nu C^\lambda]\}.
\end{eqnarray}
 One readily establishes the fact that the quantum action
 $S={\cal S}+S_{\rm GF}+S_{\rm GH}$ (1), (29), (32) is invariant under
 the complex of the background transformations (23) and the transformations
 of the ghost fields
\begin{eqnarray}
 \delta\overline{C}&=&(\partial_\mu\overline{C})\xi^\mu,\;\;\;
 \delta C=-C^\mu\partial_\mu\zeta+(\partial_\mu C)\xi^\mu,
 \nonumber\\
 \delta\overline{C}^\mu&=&-\overline{C}^\nu\partial_\nu\xi^\mu
 +(\partial_\nu\overline{C}^\mu)\xi^\nu, \\
 \delta{C}^\mu&=&-{C}^\nu\partial_\nu\xi^\mu
 +(\partial_\nu{C}^\mu)\xi^\nu.\nonumber
\end{eqnarray}
 From Eqs.~(23), (33) it follows immediately that the functional
 $Z(J,A)$ (25) is invariant under the complex of the initial gauge
 transformations (3), (5) of the background fields
 $A=(e^i_\mu,\omega_\mu)$ and the transformations of the sources
 $J=(J_i^\mu,J^\mu)$
\begin{eqnarray}
 \delta J_i^\mu&=&\varepsilon_{ij}J^{\mu j}\zeta
 -J_i^\nu\partial_\nu\xi^\mu
 +\partial_\nu(J_i^\mu\xi^\nu),\nonumber\\
 \\
 \delta J^\mu&=&-J^\nu\partial_\nu\xi^\mu+
 \partial_\nu(J^\mu\xi^\nu).\nonumber
\end{eqnarray}
\hspace*{\parindent}
 As a consequence of the invariance of
 $Z(J,A)\equiv\exp\{\frac{i}{\hbar}W(J,A)\}$ under
 (3), (5), (34) we have the invariance of the functional
 $\Gamma=\Gamma(\overline{Q},A)$, $\overline{Q}=
 (\overline{q}^i_\mu,\overline{q}_\mu)$
\[
 \Gamma(\overline{Q},A)=W(J,A)-J\overline{Q},\;\;\;
 \overline{Q}=\frac{\delta W}{\delta J},\;\;\;
 J=-\frac{\delta\Gamma}{\delta\overline{Q}}
\]
 under the complex of the transformations (3), (5) and the
 transformations of the fields $\overline{Q}$
\begin{eqnarray}
 \delta\overline{q}^i_\mu&=&\varepsilon^{ij}
 \overline{q}_{\mu j}\zeta+\overline{q}^i_\nu\partial_\mu\xi^\nu+
 (\partial_\nu\overline{q}^i_\mu)\xi^\nu,\nonumber\\
 \\
 \delta\overline{q}_\mu&=&\overline{q}_\nu\partial_\mu\xi^\nu+
 (\partial_\nu\overline{q}_\mu)\xi^\nu.\nonumber
\end{eqnarray}
 By virtue of Eq.~(35), the effective action $\Gamma(A)$ defined in the
 background field method by the rule
\[
 \Gamma(A)=\Gamma(\overline{Q},A)|_{\overline{Q}=0},
\]
 is invariant under the initial gauge transformations
 (3), (5) of the background fields $A=(e^i_\mu,\omega_\mu)$.
\section{Conclusion}
 In this paper we have considered the model$^7$ of two-dimensional gravity
 with dynamical torsion (1), (3), (5) and performed its quantization using
 different Lagrangian methods. Thus, for the model in question we have
 obtained the generating functionals of Green's functions (17), (22) within
 the BV$^{23}$ and BLT$^{24}$ quantization formalisms for general gauge
 theories as well as the analog (25), (29), (32) of the generating functional
 of Green's functions applied for the construction of the gauge-invariant
 effective action in the background field method (see, for example,
 Ref.~25). Notice that, as far as the classically equivalent$^{20}$ model
 (7), (11) with auxiliary fields is concerned, not only may it prove to be
 non-equivalent to (1), (3), (5) at the quantum level, but also its
 quantization should pose a considerable problem involving the solution of
 the generating equations for an open algebra of gauge transformations, as
 required (in contrast to the study of Ref.~[20]) by the consistent
 quantization procedure.$^{23,24}$\\

\vspace{.5cm}
\noindent
{\Large\bf Acknowledgments}\\

\noindent
 The authors are grateful to I. L. Buchbinder for useful discussions.
 The work has been partially supported by the Russian Foundation for
 Basic Research (RFBR), project 96--02--16017, as well as by the joint
 project of Deutsche Fortschungsgemeinschaft and Russian Foundation for
 Basic Research (DFG--RFBR), 96--02--00180G.
\newpage
\noindent
{\Large\bf References}\\

\begin{enumerate}
\item
R. Jackiw, {\it Nucl. Phys.} {\bf B252} (1985) 343.

\item
C. Teitelboim, {\it Phys. Lett.} {\bf B126} (1983) P. 41;
T. Banks and L. Susskind, {\it Int. J. Theor. Phys.} {\bf 23} (1984)
475; I. M. Lichtzier and S. D. Odintsov, {\it Mod. Phys. Lett.}
{\bf A6} (1991) 1953.

\item
N. Sanchez, {\it Nucl. Phys.} {\bf B266} (1986) 487.

\item
J.D. Brown, M. Henneaux and C. Teitelboim, {\it Phys. Rev.} {\bf D33}
 (1986) 319.

\item
R. Balbinot and R. Floreanini, {\it Phys. Lett.} {\bf B160}
(1985) 401; R. Floreanini, {\it Ann. Phys. (N. Y.)} {\bf 167} (1986) 317.

\item
T. Fukuyama and K. Kamimura, {\it Phys. Lett.} {\bf B160} (1985) 259.

\item
M.O. Katanaev and I.V. Volovich, {\it Ann. Phys.} {\bf 197} (1990) 1.

\item
M. Martellini, {\it Ann. Phys. (N. Y.)} {\bf 167} (1986) 437.

\item
R. Marnelius, {\it Nucl. Phys.} {\bf B211} (1983) 14.

\item
L. Brink, P. Di Vecchia and P. Howe, {\it Phys. Lett.} {\bf B65}
(1976) 471; S. Deser and B. Zumino, {\it Phys. Lett.} {\bf B65} (1976) 369.

\item
M.B. Green and J.D. Schwarz and E. Witten, {\it Superstring Theory}
(Cambridge Univ. Press, London, 1987).

\item
P. Van Nieuwenhuizen, {\it Nucl. Phys.} {\bf B211} (1983) 14.

\item
M.O. Katanaev, {\it J. Math. Phys.} {\bf 31} (1991) 2483; {\bf 34} (1993)
22.

\item
W. Kummer and D.J. Schwarz, {\it Phys. Rev.} {\bf 45} (1992) 3628.

\item
S.N. Solodukhin, {\it Int. J. Mod. Phys.} {\bf D3} (1994) 269.

\item
T. Strobl, {\it Int. J. Mat. Phys.} {\bf A8} (1993) 1383.

\item
P. Schaller and T. Strobl, {\it Class. Quant. Grav.} 1994 {\bf 11}  331.

\item
F. Haider and W. Kummer, {\it Int. J. Mod. Phys.} {\bf A9} (1994) 207.

\item
W. Kummer and D.J. Schwarz, {\it Nucl. Phys.} {\bf B382} (1992) 171.

\item
N. Ikeda and K. Izawa, {\it Prog. Theor. Phys.} {\bf 89} (1993) 223.

\item
M.O. Katanaev and I.V. Volovich, {\it Phys. Lett.} {\bf 175B} (1986)
413.

\item
P.M.Lavrov and I.V.Tyutin, {\it Yad. Fiz.} {\bf 50} (1989) 1467.

\item
I.A. Batalin and G.A. Vilkovisky, {\it Phys. Lett.} {\bf 102B}
 (1981) 27; {\it Phys. Rev.} {\bf D28} (1983) 2567.

\item
 I.A. Batalin, P.M. Lavrov and I.V. Tyutin, {\it J. Math. Phys.} {\bf 31}
 (1990) 1487;
 {\bf 32} (1991) 532; {\bf 32} (1991) 2513.

\item
I.L. Buchbinder, S.D. Odintsov and I.L. Shapiro,
{\it Effective Action in Quantum Gravity} (IOP Publ, Bristol and
Philadelphia, 1992).

\item
B.S. De Witt, {\it Dynamical Theory of Groups and Fields} (Gordon and
Breach, New York, 1965).

\item
L. F. Abbott,  {\it Nucl. Phys.} {\bf B185} (1981) 189.
\end{enumerate}
\end{document}